\documentclass[a4paper,12pt,notitlepage]{article}
\usepackage{latexsym}
\newenvironment{changemargin}[2]{\begin{list}{}{%
\setlength{\topsep}{0pt}%
\setlength{\leftmargin}{0pt}%
\setlength{\rightmargin}{0pt}%
\setlength{\listparindent}{\parindent}%
\setlength{\itemindent}{\parindent}%
\setlength{\parsep}{0pt plus 1pt}%
\addtolength{\leftmargin}{#1}%
\addtolength{\rightmargin}{#2}%
}\item }{\end{list}}
\begin{document}

\begin{changemargin}{-0.4cm}{-0.4cm}

\begin{center}
\Large
{\bf Edge-state effects on the magneto-polarizability of nanographite layers}
\end{center}
\normalsize
\begin{center}
St\a'ephane Pleutin* and Alexander Ovchinnikov\\
Max-Planck-Institut f\"ur Physik Komplexer Systeme, N\"othnitzer
Stra\ss e 38, D-01187 Dresden\\
\small
*Corresponding author. Fax:+49-351-871-21-99. E-mail address: pleutin@mpipks-dresden.mpg.de
\end{center}
\normalsize

Only very recently a new type of magnetism similar to a spin glass state was
found for activated carbon fibers prepared at heating temperatures over 1200\char23 C [1]. These materials are supposed to be well described by a disordered network of
nanographite layers with a characteristic length of some tens angstroem [2]. In connection with recent theoretical works on graphite
ribbons, it was suggested by Shibayama et al. that the spins observed in [1]
are due to the edge states of the $\pi$ electrons whose wave functions are
mainly localized on the boundaries of the layers. Indeed, several
theoretical studies of graphite ribbons [3,4], had predicted the
existence of such states. These
edge states are expected to effect novel electrical and magnetic properties such
as a paramagnetic behaviour at low temperatures [4]. These
theoretical predictions could also be of importance for disordered or
amorphous carbon materials where effective nanographite areas appear. 

The purpose of this letter is to propose a novel way to identify the existence
of edge states in activated carbon fibers or any other disordered or
amorphous carbon compounds. Indeed, we believe that the study of their
polarizability under an uniform and static magnetic field
should give a very direct way to identify such states. Similar studies have
already been proposed in [5] and [6] for cylindrical-like systems such as carbon
nanotubes, on the base of semi-classical arguments and full quantum calculations, respectively. In these articles, it was pointed out that the static
magneto-polarizability of cylindrical systems shows a very intricate behaviour
due to the Aharonov-Bohm effect. It was then suggested that the careful
analysis of these complicated features should provide a tool to get informations
about the excited states of these systems. The same kind of idea is followed
in this work.

As a first example, we consider here perfect nanographite layers with rectangular
shapes (figure 1); more complicated cases will be subject of subsequent
publications. At the boundaries two kinds of edge appear: the so-called armchair
and zig-zag edges. The $\pi$ electrons are described by an usual tight
binding model. Since we want to discuss only orbital magnetism, they are considered as spinless particles.

\begin{equation}
\label{tb}
\hat{H}=\sum_{<i,j>}t_{ij}a^{\dagger}_i a_j
\end{equation} where the operator $a^{\dagger}_i$ ($a_i$) creates
(annihilates) a $\pi$ electron on site $i$ of the nanographite, $t_{ij}$ is
the transfer integral between sites $i$ and $j$ - a real quantity without magnetic field. The summation
in (\ref{tb}) is over nearest-neighbours sites.

 As found in references [3,4] for graphite ribbons, our analysis of the
eigenstates of (\ref{tb}) for nanographite layers shows the existence of edge
states. Moreover, in agreement with the results of these previous studies,
these states have non-bonding character, are preferentially localized on the
zig-zag edges and their energies are in the middle of the gap, showing a
small dispersion. The number
of edge states depends on the size of the nanographite rectangles.

Next we apply an uniform magnetic field $\vec{B}$, perpendicular to the nanographite
layer. In the tight-binding model (\ref{tb}) we then proceed to the so-called
London substitution, $t_{ij} \rightarrow  t_{ij}e^{i2\pi \frac{e}{ch}\int^j_i
\vec{dl}\vec{A}}$, where $\vec{A}$ is the vector potential. For calculations we
choose, as in [4], the Landau gauge. 

By comparison with the work done in [5,6], one can say that the edge states define an effective
rings. We are then again in the situation described in [5,6] of a cylindrical system submitted to a magnetic field
applied along the cylindrical axis. The main idea of our proposal is the following: only the
edge states will react to a magnetic field sufficiently small that the
characteristic length of the corresponding cyclotron radius remains comparable to the system size. We
want to stress in this communication that this sensibility of the edge states
should be apparent in the behaviour of the polarizability. The analysis of this
physical quantity as function of the magnetic-field should then give a complementary
experimental tool for the study of carbon materials such as carbon fibers and
amorphous carbons.

We apply an electric field $\vec{E}$, parallel to the nanographite plane. It
could be oriented in any direction without changing qualitatively the results
shown here. Next, we calculate the polarizability,

\begin{equation}
\label{dipole}
P(T,\phi)=\frac{1}{|\vec{E}|}\frac{Tr( \hat{d}e^{-\beta(\hat{H}-\mu)})}{Tr(e^{-\beta(\hat{H}-\mu)})}
\end{equation}where, as usual, $\beta=\frac{1}{k_B T}$ ($T$ is the temperature), $\mu$ is the chemical
potential, and $\hat{d}$ is the dipole operator, $\hat{d}= e\sum_{i}
\vec{r}_{i}.\vec{E}a^{\dagger}_{i}a_{i}$. The site vector $\vec{r}_{i}$ is defined with respect to an arbitrary
origin. $\phi=2\pi\frac{e}{ch}S|\vec{B}|$ is the magnetic flux in units of $\phi_0=\frac{ch}{e}$, where $S$ is the
surface of a benzene ring.

If the size of the nanographite plan is large enough, signatures of the
edge-states should appear in its magneto-polarizability for not too large values of the magnetic field. A characteristic result is shown in figure 2 for a nanographite
layer with 78 benzene rings. In this particular case, there are two
edge states with quasi-degenerated energies at zero. An appropriate way to measure the
localisation of the different wave functions is to evaluate the sum over the
square root of their coefficients along the edges, $W_e=\sum_{i\in {\rm edge}}
a_i^2$. This quantity shows us for the chosen example that: (i) the edge
states are at 90$\%$ localized on the
edges, (ii) there is a small diffusion of the wave function from the edges to
the interior of the plane induced by the magnetic field. 

Figure 2 shows the corresponding polarizability as function of the
magnetic flux for two different fillings; a magnetic flux of $\phi/\phi_0=10^{-5}$ corresponds
approximately to a field of 1 Tesla. The full curve is for 78 electrons,
which corresponds to the half-filling case, and the dotted curve is for 77 electrons. The temperature is very small
$k_BT=10^{-5}t_{ij}$, which can be reasonably estimated to $T\simeq 0.1K$, a
quantity much smaller than the gap between the energies of the edges states and
the rest of the spectrum. So
one can conclude that the edge states are occupied only at
half-filling. Since only the polarizability for the half-filled band is
noticeably a function of the magnetic flux, we may conclude that the edge
states are responsible for the strong quadratic dependence obtained for a reasonably small
magnetic field (about a few Tesla). It should be stressed that the relative
variations in the figure 2 for the half-filled case could be detected
experimentally nowaday. To conclude, our claim is that it is possible to identify
the existence of edge states in activated carbon fibers or amorphous carbon by
measurement of
the magneto-polarizability of these materials. Moreover, we believe that informations about the effective sizes of the
graphite like regions in the carbon fibers can be gained by careful studies of
the quadratic behaviour pointed out here. More systematic studies will be published elsewhere.

Before ending this communication, one should stress first, that
electron-electron interactions should be incorporated to get better estimate of
the absolute value relative of the magneto-polarizability. However, the
screening effect should not depend on magnetic field and therefore, should not
change its relative variations (figure 2). Second, the Zeeman and spin-orbit
coupling should be also incorporated for a complete treatment. However, for
the values of the field considered here, these interactions are less important
than those discussed in this work. Thus they should not change drastically our results [6]. Finally, it may be difficult in practice to apply the magnetic field perpendicular to the plan. In this case, we should simply replace $H$ by $H \cos
\theta$ in our calculations, where $\theta$ is the angle between the magnetic field and the normal to
the plan. Moreover, for disordered materials several averaging procedures
should be applied: averaging over the size and shape of the nanographites and averaging
over $\theta$.\\
\end{changemargin}
\large
{\bf References}
\normalsize

[1] Y. Shibayama, H. Sato, T. Enoki and M. Endo, Phys. Rev. Lett. {\bf 84}, 1744
(2000).

[2] M.S. Dresselhaus, G. Dresselhaus and P. Eklund, "Science of fullerenes and
carbon nanotubes",  Academic Press (1996).

[3] M. Fujita, K. Wakabayashi, K. Nakada and K. Kusakabe,
J. Phys. Soc. Jpn. {\bf 65}, 1920 (1996).

[4] K. Wakabayashi, M. Fujita, H. Ajiki and M. Sigrist, Phys. Rev. {\bf B 59},
8271 (1999).

[5] P. Fulde and A.A. Ovchinnikov, Eur. Phys. J. {\bf B 17}, 623 (2000).

[6] S. Pleutin and A.A. Ovchinnikov, cond-mat/0108132


\end{document}